\documentstyle[epsfig,letters]{mn1}

\def\la{\mathrel{\hbox{\rlap{\hbox{\lower4pt\hbox{$\sim$}}}\hbox{$<$}}}}
\def\ga{\mathrel{\hbox{\rlap{\hbox{\lower4pt\hbox{$\sim$}}}\hbox{$>$}}}}

\title[Cluster-Cluster Lensing]{Cluster-Cluster Strong Lensing:
Expectations and Detection Methods}
\author[Cooray, Holder, \& Quashnock]{Asantha ~R. ~Cooray,
Gilbert ~P. ~Holder, Jean ~M. ~Quashnock\\
Department of Astronomy and Astrophysics, University of Chicago,
Chicago IL 60637.\\
E-mail: asante@hyde.uchicago.edu, holder@oddjob.uchicago.edu,
jmq@oddjob.uchicago.edu}
\date{\today}
\begin{document}
\maketitle

\begin{abstract}
We calculate the all-sky number of
galaxy clusters that are expected to be
gravitationally lensed by foreground massive clusters.
We describe the redshift and number distributions
of clusters using a Press-Schechter analysis,
and model the foreground lensing clusters as singular isothermal spheres.
If $\Omega_m=0.3$ and $\Omega_\Lambda=0.7$, we expect
$\sim$ 30 cluster-cluster strong lensing events that involve
foreground X-ray luminous clusters with total mass
greater than $7.5 \times 10^{14}\, h^{-1}\, {\rm M_{\sun}}$,
or X-ray luminosity $L_x (2-10\, {\rm keV})\,
\ga 8 \times 10^{44}\, h^{-2}\, {\rm ergs\, s^{-1}}$,
and background clusters with total mass greater than
$10^{14} \, h^{-1} \, {\rm M_{\sun}}$. The number expected in
an open universe with $\Omega_m = 0.3$ is less than $\sim$ 4.
Because of uncertainty in $\sigma_8$, the root-mean-square density fluctuations
in spheres of radius 8$h^{-1}$ Mpc, the exact number of
such lensing events is uncertain by a factor of about 5.
We examine methods to detect cluster-cluster lensing events
based on optical, X-ray, and Sunyaev-Zel'dovich effect observations.
\end{abstract}

\begin{keywords}
cosmology: observations  --- gravitational lensing ---
galaxies: clusters --- galaxies: clusters: individual: Cl0024+17
\end{keywords}

\section{Introduction}

It is now well known that the incidence  of gravitational lensing
depends strongly on the cosmological constant,
while the number density of galaxy clusters, especially at high redshifts,
depends mostly on the total mass density of the universe.
Therefore, cluster abundances and gravitational lensing
together constrain {\em both} the cosmological constant
and the mass density of the universe.
One may then expect the lensing of background clusters
by foreground clusters
to yield important information on these two cosmological parameters.

Clusters gravitationally lens background sources, such as galaxies,
into luminous optical arcs.
Cooray (1999; hereafter C99) has calculated
the all-sky number of lensed arcs due to foreground massive clusters.
We extend the analysis in C99 to study a unique situation
where a background cluster is strongly lensed
by a massive foreground cluster.
Such lensing events occur when the two clusters
lie along the same line of sight.
Given that the number density of clusters is small compared to that of
galaxies, we expect such cluster-cluster lensing events to be rare.
However, during the next decade several new missions
are expected to survey large portions of the sky,
over a wide range of wavelengths ranging from radio to X-ray,
making detection of a cluster-cluster lensing event possible.

Here, we calculate the all-sky number of
background galaxy clusters that are expected to be lensed by massive
foreground clusters, in different cosmological models.
We describe galaxy clusters with a
Press-Schechter analysis (Press \& Schechter 1974; hereafter PS)
normalized by the local cluster temperature function,
and model the foreground lensing clusters as singular isothermal
spheres (SIS).

In \S~2 we present our calculation
and discuss possible systematic errors affecting it.
In \S~3 we examine possible techniques designed to detect
cluster-cluster strong lensing events,
using optical, X-ray, and Sunyaev-Zel'dovich (SZ) effect observations.
In \S~4 we briefly consider the possibility that cluster Cl0024+17
has gravitationally lensed a background cluster,
and summarize our results in \S~5.
We follow the conventions that the Hubble
constant, $H_0$, is 100 $h$ km~s$^{-1}$~Mpc$^{-1}$, the present mean
density of the universe in units of the closure density is $\Omega_m$,
and the present normalized cosmological constant is $\Omega_{\Lambda}$.
In a flat universe, $\Omega_m+\Omega_{\Lambda}=1$.

\section{Expected Number of Lensed Clusters}

In order to calculate the number of background clusters
lensed by foreground galaxy clusters, we model the
lensing clusters as singular isothermal spheres (SIS), and use the
analytical  filled-beam approximation (Fukugita et al. 1992).
The SIS approximation to describe effective lensing cross section of clusters
should be valid given that the current observations only support a
small or zero core radius for the dark matter distribution within
galaxy clusters.
We refer the reader to the works of Cooray, Quashnock, \& Miller (1999)
and C99 for details of the lensing calculation,
in the context of background galaxies gravitationally lensed by
foreground galaxies or galaxy clusters, respectively. In both these studies,
 the redshift and magnitude distributions of optical sources
observed in the Hubble Deep Field (Williams et al. 1996)
is taken to describe the background objects.
Unfortunately, we do not have a comparable understanding
of the distribution of background clusters,
since the observational
data on the density of high redshift clusters are limited; here,
we use the PS theory to calculate the redshift distribution.

Because clusters can be studied at various wavelengths,
via optical, X-ray and radio (e.g., SZ effect) observations,
we ignore the effects of ``magnification bias'' in the present calculation.
This systematic effect, which arises in flux- or magnitude-limited surveys,
is wavelength dependent. While our present calculation
assumes a 100\% efficiency in detecting cluster-cluster lensing events,
it must be corrected to include magnification bias
for each particular observational program depending on the technique used
to recover cluster-cluster lensing events.

\subsection{Calculation method}

We derive the number density of galaxy clusters, $dn(M,z)$, between
mass range $(M,M+dM)$, from a  PS analysis:
\begin{equation}
{\frac{dn(M,z)}{dM}} =
\end{equation}
\begin{eqnarray*}
- \sqrt{\frac{2}{\pi}} \frac{\bar{\rho}(z)}{M}\frac{d\sigma(M,z)}{dM}
\frac{\delta_{c}}{\sigma^2(M,z)}
\exp{\left[\frac{-\delta_{c}^2}{2 \sigma^2(M,z)}\right]}\, , \nonumber
\end{eqnarray*}
where $\bar{\rho}(z)$ is the mean background density at redshift $z$,
 $\sigma^2(M,z)$ is the variance of the fluctuation spectrum averaged over
a mass scale $M$, and  $\delta_{c}$  is the
linear over-density of a perturbation which has collapsed and virialized.
Following Viana \& Liddle (1996), we take
\begin{equation}
\sigma(R,z)=\sigma _8(z) \Bigl( { R \over 8 h^{-1} {\rm
Mpc}}\Bigr)^{-\gamma(R)} ,
\end{equation}
where
\begin{equation}
\gamma(R) = (0.3\Gamma + 0.2)\left[2.92 + \log_{10}
\Bigl({R \over 8 h^{-1} {\rm Mpc}}\Bigr)\right] \, .
\end{equation}
The comoving radius $R$ is the radius which contains mass $M$
at the current epoch, while $\Gamma$ is the usual CDM shape parameter,
taken to be 0.23. Our presented results are
fairly insensitive to the exact choice of $\Gamma$.

In an Einstein-de Sitter universe ($\Omega_m\!=1,\Omega_{\Lambda}\!=0$),
$\sigma_8 \propto (1+z)^{-1}$.
Following Carroll, Press \& Turner (1992), we can
express growth in alternate cosmologies by a growth suppression
factor, which is approximately given by
\begin{equation}
g(\Omega_m) = {5 \over 2} \Omega_m
\left[ \Omega_m^{4/7} - \Omega_{\Lambda} +
\Bigl(1 + {\Omega_m \over 2} \Bigr)
\Bigl(1 + {\Omega_{\Lambda} \over 70} \Bigr) \right]^{-1} \, .
\end{equation}
In this notation, we can now express the growth law as
\begin{equation}
\sigma_8(z) = {\sigma_8(0)\over 1+z}\,\,
{g(\Omega_m(z)) \over g(\Omega_m(0))} \, ,
\end{equation}
where we now use the determination of Viana
\& Liddle (1998) for $\sigma_8(0)$:
\begin{equation}
\sigma_8(0) = \left\{ \begin{array}{cc}
  0.56 \, \Omega_m^{-0.34} & {\rm (open)} \\
  0.56 \, \Omega_m^{-0.47} & {\rm (flat)} \\
 \end{array} \right.
\end{equation}

The probability of strong lensing depends on both the number density and
typical mass of galaxy clusters.
For SIS models, this overall factor for the lensing probability
is given by the dimensionless parameter (Turner, Ostriker \& Gott 1984):
\begin{equation}
F\equiv 16\pi^3nR_0^3\left(\sigma_{\rm vel}\over{c}\right)^4\; ,
\end{equation}
where $n$ is the number density of galaxy clusters, $R_0\equiv
c/H_0$, and $\sigma_{\rm vel}$ is the velocity dispersion.
In order to calculate the parameter $F$, we use the observationally
determined scaling
relation between $\sigma_{\rm vel}$
and the cluster temperature, $T$ (Girardi et al. 1996),
\begin{equation}
\sigma_{\rm vel}(T) = 10^{2.56 \pm 0.03} \times
\left( \frac{T}{\rm keV}\right)^{0.56 \pm 0.05}\; ,
\end{equation}
and the relation between $T$ and cluster mass $M$ (Bartlett 1997),
\begin{equation}
T(M,z) = 6.4 h^{2/3} \left(\frac{M}{10^{15}\, M_{\sun}}\right)^{\frac{2}{3}}
(1+z)\, {\rm keV}\; ,
\end{equation}
to derive a relation between $\sigma_{\rm vel}$ and cluster mass $M$,
$\sigma_{\rm vel}(M)$.
Finally, we can write the parameter $F$ as a function of
the lens redshift, $z_l$:
\begin{equation}
F(z_l)=
\frac{16 \pi^3}{c H_0^3} \int_{M_{\rm min}}^{\infty}
\sigma_{\rm vel}(M')^4 \frac{dn(M',z_l)}{dM'}\, dM'\;.
\end{equation}
In order to calculate the number of lensed clusters, we
compute $\langle F(z_l) \rangle$, the mean value of $F(z_l)$,
which is obtained by weighting the above integral
over the redshift distribution of galaxy clusters derived
using the PS theory.
In order to calculate the number of background lensed clusters
towards a sample of X-ray selected clusters with luminosity $L_x$,
we also need a relation between $M$ and $L_x$.
We obtain this relation by combining the observed $L_x-T$ relation,
recently derived by Arnaud \& Evrard (1998),
\begin{equation}
L_x = 10^{45.06 \pm 0.03} \times
\left(\frac{T}{6\, {\rm kev}}\right)^{2.88\pm0.15}\,
0.25 h^{-2}\, {\rm ergs\, s^{-1}}\; ,
\end{equation}
and the $M-T$ relation in Eq.\ 11.
Combining $\sigma_{\rm vel}(M)$
and $dn(M,z)/dM$,
and using a minimum mass $M_{\rm min}$ of
$7.5 \times 10^{14}\, M_{\sun}$, corresponding to $L_{\rm min}$ of $8
\times 10^{44}\, h^{-2}\, {\rm ergs\, s^{-1}}$ in the 2-10 keV
band, we find  $\langle F(z_l) \rangle$ to range from
$\sim$ $4 \times 10^{-6}$, when $\Omega_m=1$,
to $\sim$ $7 \times 10^{-4}$, when $\Omega_m=0.1$.
This strong variation results from the combined dependences
of the lensing probability and the cluster density evolution on the
cosmological parameters.

In this calculation and in our results below, we have taken
the background clusters to be point sources, since Eq. 7
gives the probability of lensing solely in terms of the Einstein radius
of the lensing cluster. However, background galaxy clusters have
finite angular size, and this could have an effect
on our calculation, since {\em partial} overlap of
the background cluster with the Einstein radius of the foreground galaxy
might still constitute and be defined as a cluster-cluster lensing event.
Unfortunately, the exact amplitude of this effect is uncertain,
since it depends on what we define as a lensing event:
Do we require that half or more of the constituent emission
of the background cluster be inside the Einstein radius?
What fraction of the background cluster should be lensed in order
to constitute a detectable lensing event?

Because of these ambiguities and the additional complications
that they introduce in the computation, we have not included them here.
Nevertheless, we have estimated the approximate amplitude of
the finite-angular-size effect by considering the typical overlap
between a lensed background cluster ($z\sim 1.5$)
and the Einstein radius of a lensing foreground galaxy ($z_l\sim 0.4$),
where an overlap is defined as meaning that more than half
of the emission of the background cluster is inside the Einstein radius.
We have examined this effect for different values of
$\Omega_m$ in a flat universe, thus including the dependence of the
angular diameter distance on the mean mass density.

We find that the net effect is to increase the rate of lensing,
and hence the all-sky number of cluster-cluster lensing events,
by about a factor of two (ranging from 1.8 in a low-density universe
to about 2.2 in an Einstein-de Sitter universe).
This is small compared to other systematic uncertainties
in our calculation (see Fig.~1 below); furthermore, the 20\%
change in the factor is small compared to the overall dependence
of the number of lensed clusters on the cosmological parameters.
Thus, our main conclusions are not significantly affected
by ignoring the finite angular size of background clusters.

\subsection{Results}

Using cluster abundances at high redshift, we have calculated the expected
all-sky number of background clusters that are
gravitationally lensed by foreground clusters.
In Figure~1, we show our results
as a function of $\Omega_m$ for flat ($\Omega_m+\Omega_\Lambda=1$; {\it
solid lines}) and open ($\Omega_\Lambda=0$; {\it dashed lines})
cosmologies. The two curves for each of these cases
bound the 95\% confidence range in the
expected number of cluster-cluster lensing events based on the uncertainties
associated with $\sigma_8(0)$ and scaling relations (see, \S~2.3).

In all cases,
foreground lensing clusters
are taken to have  X-ray luminosities greater than $8 \times
10^{44}\, h^{-2}\, {\rm ergs\, s^{-1}}$ in the 2 to 10 keV band.
There are expected to be about 8000 $\pm$ 1000 such
clusters on the whole sky (Bartelmann et al. 1997).

As shown in Fig.~1, the all-sky number of lensed background clusters is
not very large:
Anywhere from about 0.7 to 60 lensed clusters with mass greater than
$10^{14}\, h^{-1}\, M_{\sun}$ are
expected in a universe with $\Omega_m=0.3$, with a higher
number in a flat universe than in an open universe.

The expected all-sky number of lensed clusters is significantly greater
than unity only in flat universes with $\Omega_m\la 0.5$
or in open universes with $\Omega_m \la 0.3$.
Even though the lensing rate for background clusters is small,
the detection of even a single cluster-cluster lensing event
provides strong evidence against a high-density universe
($\Omega_m\sim 1$), since in such a universe the expected number
is essentially zero.

If $\Omega_m$ is known {\em a priori} and is
fairly low, the effect of the cosmological constant is noticeable,
and the cluster-cluster lensing rate
could be a useful tool for differentiating
between a flat universe with a cosmological constant and an open one.
At this point, uncertainties in the predictions are dominated by the
uncertainty in $\sigma_8$. Improvements in this determination will
strongly increase the power of this test.

In general there are at most of order 0.005 to 0.05
lensed background clusters per foreground cluster.
In contrast, there are 0.2 to 0.4 gravitationally lensed arcs
(with amplification greater than 10) per foreground cluster,
with the same criteria for lensing, down to a R band magnitude of $\sim$ 21.5.
Thus, the cluster-cluster lensing rate is considerably smaller, and
a survey of area $\ga 1000$~deg$^2$ will be required to find such an event,
if we happen to live in a low-density universe with $\Omega_m\la 0.6$.

From the redshift distribution of background clusters,
as given by the PS analysis, we find that most of the lensed clusters are
at redshifts between 1.0 and 1.6, while the foreground lensing
clusters are at redshifts between 0.2 and 0.6.
This is important when considering observational programs designed
to find cluster-cluster lensing events, given the
sensitivities of current and planned surveys (see \S\ 3 below).

\begin{figure}
\psfig{file=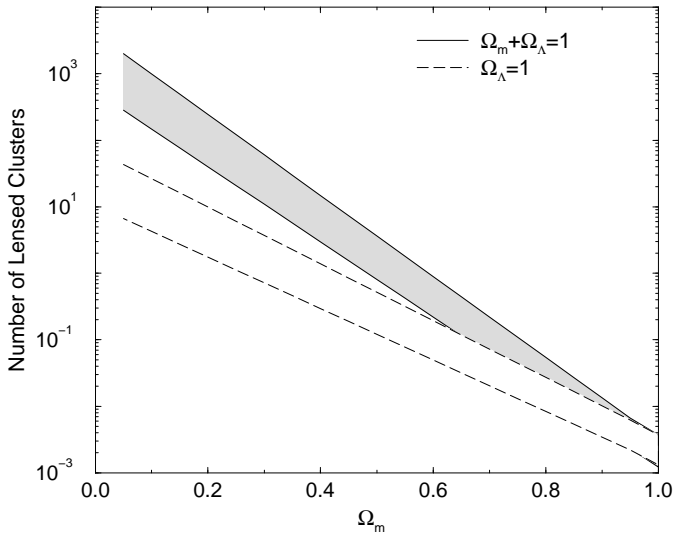,width=2.8in,angle=-90}
\caption{Expected number of all sky cluster-cluster strong lensing
events involving background clusters of masses greater than $10^{14}\,
h^{-1}\, M_{\sun}$ and foreground clusters with masses greater
than $7.5 \times 10^{14}\, h^{-1}\, M_{\sun}$ as a function of
$\Omega_m$. The solid lines are the expected numbers in a flat
cosmological model ($\Omega_m+\Omega_\Lambda=1$), while the dashed lines
are the number in a open cosmological model with
$\Omega_\Lambda=0$. The two curves represent the 95\% confidence on
the prediction based on the uncertainty in the
determination of $\sigma_8(0)$ by Viana \& Liddle (1998), and uncertainties
in the scaling relations used in the calculation (see Sect. 2.3).}
\end{figure}

\subsection{Uncertainties}

A major source of statistical
uncertainty in the calculation
comes from the accuracy of $\sigma_8(0)$ as
determined by Viana \& Liddle (1998):
The relative uncertainty in $\sigma_8(0)$ (95\% confidence)
is quoted as $+20\Omega_m^{0.1\log_{10}\Omega_m}$ \% and
$-18\Omega_m^{0.1\log_{10}\Omega_m}$ \% for the open case,
and $+20\Omega_m^{0.2\log_{10}\Omega_m}$ \% and
$-18\Omega_m^{0.2\log_{10}\Omega_m}$ \% for the flat case.
This uncertainty in $\sigma_8(0)$ dominates that of the scaling relations
as determined based on observed data. We have included the uncertainty in
$\sigma_8(0)$ and that of scaling relations, to be complete,
in our calculation to determine their effects on our prediction.
As shown in Fig.~1, the uncertainty
in $\sigma_8(0)$, as well as the ones involved with the scaling
relations, suggest that one may not be able to distinguish between
high $\Omega_m$ universes, whether spatially flat or open.
For low $\Omega_m$ values, the uncertainties are still high, but
reliable statistics on the cluster-cluster lensing
events are able to distinguish between open and flat cosmologies,
assuming $\Omega_m$ is roughly known.

For this calculation we have used the
most conservative estimate on $\sigma_8(0)$
and its uncertainty available from the literature.
It is likely that with
better statistics on cluster abundances at high redshift,
$\sigma_8(0)$ will be known more accurately in the future.
If that is the case, cluster-cluster lensing events could be a strong
probe of the existence of a cosmological constant.

A possible systematic effect that can affect the
large dependence of the predicted numbers on the cosmological
parameters comes from the
effect of $\Omega_\Lambda$  on cluster formation:
Clusters may be less
compact in a universe with a non-zero cosmological constant
(Bartelmann et al. 1997).
Such effects can only be explored with numerical simulations.
Based on numerical simulations performed by the Virgo Consortium,
Thomas et al. (1998) studied a series of clusters in four different
cosmologies, including an open model with $\Omega_m=0.3$ and
a flat model with a cosmological constant of $\Omega_\Lambda=0.7$.
The authors concluded that clusters do not exhibit differences
between open and flat cosmologies with and without
a cosmological constant and that cluster structures
cannot be used to discriminate between the two possibilities.
If Thomas et al. (1998) are correct, then
the inclusion of a cosmological constant is not
expected to change cluster mass
profiles to an extent that would affect the gravitational lensing rate.
In any case, such systematic effects are unlikely to
be nearly as large as the current uncertainty in $\sigma_8$.

\section{Detecting Gravitationally Lensed Background Clusters}

{\em Optical Data}---Since most of the studies of foreground galaxy
cluster lenses are made using optical images,
it may be possible to find cluster-cluster lensing
events by analyzing the gravitational lensing data
around known foreground cluster lenses. A lensing model of the
observed distribution of arcs can be used to estimate the redshifts of
the background galaxies.  The number density of lensed
galaxies and their optical properties, such as clustering properties
and number counts, may provide additional information needed to discern
the possible clustering of the lensed galaxies as would
be expected from a background cluster.
The properties of such galaxies should
be similar to those of cluster member galaxies, among
which effects such as the Butcher-Oemler effect
may be important. Spectroscopic observations of background lensed galaxies
can then be used to determine their exact redshifts, and to recover the galaxy
distribution within the background clusters.

In the near future,
the Sloan Digital Sky Survey (SDSS) will take both imaging and spectroscopic
data over a quarter of the sky. It is likely that the SDSS will image
most of the foreground lensing clusters within this sky area.
However, given that lensed background clusters are expected
to be at redshifts greater than 1,
and considering the resolution of SDSS observations and
the  complexity of cluster-cluster lensing events, it is
not likely that direct detection of a cluster-cluster lensing event will
be possible with the imaging data alone. The SDSS spectroscopic
survey is not expected to observe galaxies
at redshifts greater than $\sim$ 0.2; thus, analysis of the 5 color data
will be required in deducing the photometric redshifts of
foreground lensing clusters.
The same photometric data should allow the redshift
determination of background lensed sources, but these
may be limited to high magnification arcs. Thus optical
follow-up data may be necessary to identify cluster-cluster lensing events.
However, the SDSS can be used as a starting point to select candidate
cluster-cluster lensing events, which could be followed up at higher
resolution with other telescopes, and at other wavelengths.

{\em X-ray Data}---At X-ray wavelengths, one may detect
redshifted Fe lines associated with the background cluster X-ray emission,
which would be amplified by passing through the gravitational
potential of the foreground cluster.
Unfortunately, current instruments aboard  X-ray satellites
are not sensitive to emission from galaxy clusters
at redshifts greater than $\sim$ 1,
because of the $(1+z)^{-3}$ fall off of the
thermal bremsstrahlung flux at redshift $z > 1$.
Therefore, it is unlikely that cluster-cluster lensing events
can be detected with current cluster observations,
unless the background cluster is significantly magnified.

This situation will soon change with instruments aboard the {\it
Chandra} X-ray Observatory (formerly AXAF),
in particular the CCD Imaging Spectrometer (ACIS),
which can image an area of $16.9 \times 16.9$ arcmin$^2$, in the
0.2 to 10 keV band, with a point spread function of $0\farcs5$ (FWHM).
The ACIS allows detection of point sources down to $\sim 2 \times 10^{-15}$
ergs cm$^{-2}$ s$^{-1}$ in an image with exposure time of 10$^5$ seconds
at the 5$\sigma$ level\footnote{AXAF Observatory Guide 1997, AXAF Science
Center, available at http://asc.harvard.edu/USG/docs/docs.html}.
The CCDs have an intrinsic energy resolution $E/\Delta E$ of $\sim$ 5 to 50
across the energy band, and when combined with the High Energy
Transmission Gratings (HETG), the ACIS spectroscopic arrays are expected
to provide resolutions, $E/\Delta E$, as high as 1000.
For example, the planned 100 ksec GTO observation of lensing
by the well-known cluster A370 can be used to detect
any lensed cluster along the line of sight out to at least a redshift
of $\sim$ 1.5, ignoring any amplification of such a cluster due to
gravitational lensing. However, even if such background emission exists,
it may very well be confused
with emission from the foreground cluster.
The best approach, then, is to look for redshifted lines, such as
the Fe K-line, in the spectroscopic data.\footnote{While preparing this work,
we learned of a proposal to search, with ASCA (Watanabe et al. 1997),
for a redshifted Fe K-line that would be a signature of
a background cluster behind Cl0024+17.}

{\em SZ Data}---Unlike the X-ray, where the intensity detected decreases
with increase in
redshift, the SZ effect (Sunyaev \& Zel'dovich 1980)
in principle can detect, independent of redshift,
background clusters that have been lensed by a foreground cluster.
Because the X-ray emission falls
off as $(1+z)^{-3}$ and the SZ effect does not strongly depend
on redshift, the total
CMBR temperature change due to the SZ effect along the line of sight of two
clusters at different redshifts is different from what is
inferred from the total
X-ray emission observed alone.
Thus, combining the SZ data with X-ray observations
provides a strong tool to recover background
clusters. An efficient method to recover cluster-cluster
lensing events is to study X-ray/SZ correlations or
cross-correlate a large area SZ survey with
X-ray observations, to find clusters that deviate from
such relations. The SZ cluster catalog
that is expected to be produced by Planck and the large area X-ray
survey with XMM may afford us with such a possibility.

\section{Cl0024+27: Possible Cluster-Cluster Lensing Event?}

The redshift distribution of the lensed galaxies behind Cl0024+17 has been
estimated by Fort, Mellier \& Dantel-Fort (1997).
The analysis by Fort et al. (1997) suggests two groups of sources,
one with 60\% $\pm$ 10\% of background sources
at redshifts between 0.9 and 1.1 and another with
40\% $\pm$ 10\% of the background sources at redshifts $\sim$ 3.
The population at $z \sim$ 1 is more interesting:
Could it consist of a background cluster that has been
lensed by Cl0024+17?

Cl0024+17 is being observed interferometrically
with the BIMA array at 28.5 GHz,
as part of an SZ cluster survey (e.g., Carlstrom, Joy, \& Grego 1996).
A comparison of these recent
interferometric SZ images with X-ray data may yield
important information on the cluster gas properties of Cl0024+17.
These inferred cluster gas properties
could appear significantly different than those of most other clusters,
if there is a lensed background cluster behind Cl0024+17.

If Cl0024+17 has lensed a background cluster,
then the mass and redshift of this lensed cluster
can be used to constrain the cosmological parameters;
unfortunately, both these quantities are not currently known,
so we leave this task to be carried out in a future paper.

\section{Summary}

Using the redshift distribution of background galaxy clusters obtained using
the Press-Schechter formalism, we have calculated the expected all-sky
number of cluster-cluster lensing events. The
number in an open cosmology with $\Omega_m=0.3$
is between 1 and 10, for background clusters with total masses greater than
$10^{14} \, h^{-1} \, M_{\sun}$, with the expected number counts increasing
to be between 10 and 100 for the corresponding low-density flat universe.
In general, a significantly higher number
is expected in a universe with a cosmological constant, with this difference
becoming larger at lower matter densities.

Lensed background clusters may be found in optical
wavelengths from an analysis of lensing and spectroscopic data.
At X-ray wavelengths, spectroscopic observations can be
used to observe redshifted Fe lines from the background cluster.
The most efficient method to find these rare cluster-cluster
events is to correlate a large area SZ survey of clusters, such
as from Planck, with a similar survey in X-ray wavelengths, such
as from XMM.
The mass and redshift of such a lensed cluster can be used to constrain
the cosmological parameters.

\section*{Acknowledgments}

We would like to acknowledge useful discussions with John Carlstrom,
and Cole Miller. We would also like to thank the referee, Yannick Mellier, for
detailed comments on our paper. This work was supported in part by
NASA grant NAG~5-4406 and NSF grant DMS~97-09696 (JMQ),
and by NASA grant NAG~5-2788 and the DOE (GPH).
ARC acknowledges partial support from the
McCormick Fellowship at the University of Chicago,
and a Grant-In-Aid of Research from the National Academy of
Sciences, awarded through Sigma Xi, the Scientific Research Society.

\end{document}